\documentclass[graybox]{svmult}

\usepackage{type1cm}        
\usepackage{makeidx}         
\usepackage{graphicx}        
\usepackage{multicol}        
\usepackage[bottom]{footmisc}

\usepackage{newtxtext}       %
\usepackage{newtxmath}       

\makeindex             

\begin{document}

\title*{Mobility driven Cloud-Fog-Edge Framework for Location-aware Services: A
Comprehensive Review}
\titlerunning{Review of Mobility driven Cloud-Fog-Edge Framework} 
\author{Shreya Ghosh and Soumya K. Ghosh}
\institute{Shreya Ghosh \at Department of Computer Science and Engineering Indian Institute of Technology Kharagpur, \email{shreya.cst@gmail.com}
\and Soumya K. Ghosh \at Department of Computer Science and Engineering Indian Institute of Technology Kharagpur \email{skg@cse.iitkgp.ac.in}}
%
%
\maketitle

\abstract*{With the pervasiveness of IoT devices, smart-phones and improvement of location-tracking technologies, huge volume of heterogeneous geo-tagged (location specific) data is generated facilitating several location-aware services. The analytics with this spatio-temporal (having location and time dimensions) datasets provide varied important services such as, smart transportation, emergency services (health-care, national defence or urban planning). While cloud paradigm is suitable for the capability of storage and computation, the major bottleneck is network connectivity loss. In time-critical application, where real-time response is required for emergency service-provisioning, such connectivity issues increases the latency and thus affects the overall quality of system (QoS). To overcome the issue, fog/ edge topology is emerged, where partial computation is carried out in the edge of the network to reduce the delay in communication. Such fog/ edge based system complements the cloud technology and extends the features of the system. This chapter discusses cloud-fog-edge based hierarchical collaborative framework, where several components are deployed to improve the QoS. On the other side. mobility is another critical factor to enhance the efficacy of such location-aware service provisioning. Therefore, this chapter discusses the concerns and challenges associated with mobility-driven cloud-fog-edge based framework to provide several location-aware services to the end-users efficiently.}

\abstract{With the pervasiveness of IoT devices, smart-phones and improvement of location-tracking technologies, huge volume of heterogeneous geo-tagged (location specific) data is generated facilitating several location-aware services. The analytics with this spatio-temporal (having location and time dimensions) datasets provide varied important services such as, smart transportation, emergency services (health-care, national defence or urban planning). While cloud paradigm is suitable for the capability of storage and computation, the major bottleneck is network connectivity loss. In time-critical application, where real-time response is required for emergency service-provisioning, such connectivity issues increases the latency and thus affects the overall quality of system (QoS). To overcome the issue, fog/ edge topology is emerged, where partial computation is carried out in the edge of the network to reduce the delay in communication. Such fog/ edge based system complements the cloud technology and extends the features of the system. This chapter discusses cloud-fog-edge based hierarchical collaborative framework, where several components are deployed to improve the QoS. On the other side. mobility is another critical factor to enhance the efficacy of such location-aware service provisioning. Therefore, this chapter discusses the concerns and challenges associated with mobility-driven cloud-fog-edge based framework to provide several location-aware services to the end-users efficiently.}
\section{Introduction}
\label{sec:1}
With the rapid development of sensor and communication technologies, GPS equipped devices and Internet
of Things (IoT), varied objects such as people, resources, vehicles are interconnected and intertwined in
anywhere at any time. Alongside, with the proliferation of mobile phone users and deployment of
GPS enabled smart-devices, a huge amount of GPS traces of different geographical regions are easily available.
This massive amount of GPS traces has fostered various research directions namely human movement
behavior or activity learning \cite{thump},\cite{ctg},\cite{activity} traffic analysis, improved route planning\cite{icacni}, \cite{sinha2018hybrid} and resource allocation \cite{overview} -
which subsequently lead to smart-living of people. While IoT provides seamless connectivity to correlate
people and objects, cloud paradigm offers distributed platform to efficiently carry out the compute-intensive
tasks. Furthermore, with the latest technology, smart mobile devices are emerging as varied application
enablers for customized users’ recommendation systems, e-health apps and intelligent route planner. Mobile
cloud computing (MCC) promotes innovative solutions and approaches to leverage the computational and
storage power of cloud computing and extend the applications and services to mobile phone users on
demand basis. In recent times, IoST (Internet of Spatial Things) \cite{iost} has been emerged, which integrates IoT
and spatio-temporal data. The analysis of spatio-temporal traces such as movement information, traffic data,
weather information, help to incorporate context, and thus adds more intelligence in the processing. 

\begin{figure*}
\centering
\includegraphics[scale = 0.5]{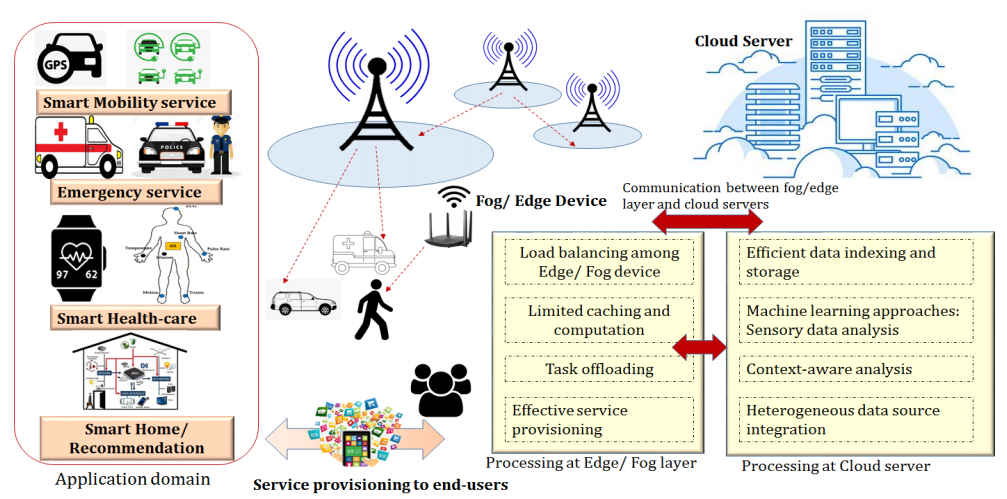}
  \caption{Overall Architecture of the mobility-aware cloud-fog-edge network}
  \label{fig1} 
\end{figure*}
\par In this direction, this chapter focuses on several aspects of mobility-aware cloud-fog-edge computing and we
put forward the future research avenues and open challenges in this research domain.
Fig. 1 illustrates the overall architecture of the mobility-aware cloud-fog-edge network. As depicted, there
are several applications, namely smart transportation system, smart mobility services, smart home etc. It
may be noted that when the user is in move, the seamless connectivity becomes a challenging issue which in
turn increases the delay/ service-provisioning time and affects the QoS. The subsequent increase in delay of
delivering result may be fatal in case of emergency services such as ambulance or fire extinguisher car. On
the other hand, mobility related information plays an important role \cite{mobiiost}. If the optimal route (less congestion
and distance) can be extracted apriori, then the service provisioning time can be reduced further. In this direction, the framework has three layers. In the bottom layer, the end-users are present. In the top-most
layer, the compute-intensive tasks such as mobility-analysis, sensor-data mining and health-data analysis
are carried out in distant cloud servers. The intermediate layers (Fog and edge layer) are used to cache
processed information and communicate with the agents for fast delivery of the service.
This chapter aims to provide a systematic survey focusing on different research aspects and existing works
in all these layers (cloud-fog-edge). Furthermore, the chapter also highlights the challenging and interesting
applications using this cloud/fog/edge networks. The major contributions of this chapter are summarized
as follows:
\begin{enumerate}
    \item A novel taxonomy based on the existing approaches and algorithms to provide mobility-aware services in
the cloud-fog-edge hierarchical network is presented.
\item The data processing and machine learning techniques at fog/edge and cloud servers are systemically discussed.
\item A topology of varied applications and services provisioned by the hierarchical network is presented.
\item The open research challenges and issues are discussed to provision intelligent and efficient location-based services.
\end{enumerate}
This survey will be beneficial for researchers, policymakers and can act as the
foundation of mobility-driven cloud-fog-edge network.
\section{Motivations and related computing paradigms}
In this section, we explain the motivation or utility of mobility-aware cloud-fog-edge framework, and briefly describe the definitions of cloud, fog, edge nodes and mobility-modelling. We also refer few use-cases where mobility-driven framework is necessary. 
\par Owing to the huge amount of data generated from varied IoT devices, it is crucial to store, manage and analyse for extracting meaningful information from the datasets. Few examples of the datasets are movement data from vehicles, people; climatology parameters from the sensor nodes; or data collected from smart-home system. Cloud computing is the on-demand availability of computer system resources, and facilitates services over the Internet. It may be noted that most of the tech giants host cloud services and provides public cloud platforms, such as, Google Cloud Platform (GCP), Amazon EC2, Microsoft Azure, IBM Bluemix etc. In summary, this technology is beneficial for its flexibility, efficiency and on-demand service. The cloud servers or data-centers help to store this huge amount of historical records to analyse and find patterns. This in turn helps to facilitate applications such as smart and effective transportation, intelligent defence techniques or weather prediction. In addition, most of the times the accumulated data from the IoT or sensors are unstructured, and partial computation is required near the source of the data. Again, the IoT devices or sensors span a large geographical area, and sending data to distant cloud servers frequently affects the efficacy of the system as a whole. This gap is managed by the fog/edge computing. The fog/ edge computing brings down the computing closer to the end-user or the devices where data has been collected, unlike carrying out all computations in the cloud data-centers. The users can get the storage or computing services at the edge of the network using this cloud-fog-edge collaborative framework. 
\par Any network device with
the capability of storage, computing, and connectivity can be used as fog/ edge nodes. For instance, the routers, switches, video surveillance cameras etc. deployed at any location with a network connection. These nodes accumulate data and can partially compute, if required. The data processing is performed at the edge network \cite{edge}, which consists of end devices, such as, 
mobile phone, border routers, bridges, set-top boxes, base stations, wireless access points etc. It may be noted that these must have necessary capabilities for supporting edge computation. In summary, edge computing provides faster responses, and also reduces the need of sending bulk data to the cloud datacenters. Integrating the edge and cloud paradigms, several new research topics have been emerged. Mobile Edge computing (MEC) and Mobile Cloud computing (MCC) are two prominent research areas in this domain. MEC is one of the key enablers of smart cellular base stations. It combines the capability of edge servers alongwith the cellular base
stations \cite{mec}. The connection with the distant cloud server is optional in MEC. Moreover, researchers are working such that MEC can support 5G communication.  
In short, mobile edge computing aims to provide faster cellular services for the customers and thus, enhances network efficiency. On the other hand, now-a-days people tend to execute necessary tasks/ application in their handheld devices. But these handheld devices are resource-hungry and have limited storage and computation capability. Hence, it is better to perform or offload compute intensive tasks outside the handheld devices. In such scenarios, mobile cloud computing plays an important role. The light-weight cloud servers \textit{cloudlet} \cite{cloudlet} are placed at the edge network. Like MEC, MCC combines the capability and features of cloud computing, mobile computing and wireless communication for better Quality of Experience (QoE) of the end-users. 
\par However, there are few challenges in this collaborative framework. Firstly, how we can manage the fog/ edge computing infrastructure and what resource allocation scheme can be adapted. Since, these fog/ edge nodes have limited resources, proper resource management should be adapted when large number of service-requests are made at a particular instance. Moreover, several factors such as service availability, power or energy consumption\cite{das2018rescue}, latency or delay should be considered while developing such framework. Therefore, the mapping of cloud/ fog/ edge nodes to several applications remains a challenging issue. Another critical part is security and privacy issues such as, trust management, access control etc. A proper security model based on the sensitivity of the datasets and requirements of the application is much needed in such cloud/ fog/ edge infrastructure. 
\par There are several use-cases which utilize such cloud-fog-edge collaborative framework \cite{overcc},\cite{clawer}. The work \textit{Mobi-IoST} (Mobility-aware Internet of Spatial Things) \cite{mobiiost} illustrates an example of time-critical application, where latency or delay is very important and can be fatal. For instance, in an ambulance, continuous monitoring of patient's vital health parameters such as blood-pressure, pulse-rate, body-temperature etc. is required. These data are collected by IoT devices and the accumulated health data is sent to the nearby fog device through a client application. In this work, authors have used Road Side Units (RSU) as fog device, while the moving agent is the edge device. The preliminary checking of the health data is carried out in RSUs and in case any abnormality is detected, the data is sent to the cloud server. The cloud datacenter extracts the location of the ambulance, and redirects it towards the nearest health center. In the paper, authors have also emphasized the present state of the traffic is important, since the ambulance or any vehicle with patient needs to travel the roads with minimum congestion. Another work, named \textit{Locator}\cite{locator},  develops a hierarchical framework with cloud, edge, fog nodes to provide food delivery services in minimum delay. There are also several sub-domains of IoT, such as, Internet of Multimedia Things (IoMT), Internet of Health
Things (IoHT) \cite{mobnew}, Internet of Vehicles (IoV) etc. \cite{iost}. The work focusing on Internet of Health Thing using delay-aware fog network is mentioned in \cite{fogioht}. It may be noted that mobility or continuous change of locations
of users or agents is a challenging, since connectivity may be lost. Therefore, analysing the movement patterns of users is important to enhance the quality of service. Mobility analytics is an integral part of developing an effective and delay-aware solution for any mission-critical or time-critical application.
\par In brief, after analysing the features of cloud-fog-edge computing, the challenges in this domain can be listed as follows.
\begin{itemize}
    \item \textbf{Resource Management}: Since the fog and edge nodes have limited resources, it is difficult to assign large scale analytics in resource constrained nodes. Therefore, proper resource management modules should be developed to avoid this bottleneck. One aspect is adapting distributed fog/ edge environment to cope up with the growing data amount. Also there must be specific policies to assign computational tasks and services among edge, fog and cloud nodes. Data visualization through web-interfaces are also not easy task through edge or fog nodes.
    \item \textbf{Mobility Sensitive}: Seamless connectivity due to the mobility of IoT devices
is a critical concern for time-critical applications. The connection interruption and consequently the increase in
delay affects the QoS. With the rapid use of mobile (smart) devices, the framework must be able to accommodate mobility data. How to analyse the huge amount of movement information and extract useful patterns are challenging tasks. Moreover, predicting next location-sequences of agents is also crucial, since it may help to take decision or offload task in the nearby fog nodes. However, there are several factors such as, present traffic condition, users' own preferences and time of the movement etc. These external contexts make the location prediction task more difficult. On the other hand, mobility data is sensitive, and proper measures must be taken to secure the whereabouts of the users. 
    \item \textbf{Security Aspect}: Fog or edge nodes are highly vulnerable to security attacks. Since these nodes handle and manage sensitive data (health related or mission critical, like Defence application), proper security measurements are must. Access control schemes, including, authenticated access to services and nodes, security algorithms are required in distributed paradigm like edge/ Fog computing are hard to ensure. Again, strict implementation or methods of security mechanisms affect the quality of service of edge and fog computing. 
    \item \textbf{Infrastructure or Organizational structure}:  As mentioned, the framework has different types of components, like cloud servers, edge and fog nodes. Several objects like routers, access points can act as potential edge and fog computing infrastructure. Therefore, the processors of these components are quite different. Implement an end-to-end framework using different components is a really difficult task. It is absolutely necessary to select suitable devices based on on operational requirements and execution environment. The resource configuration and location of the deployment are also two major factors to provide better service.

\end{itemize}
In the next part, we will discuss the taxonomy in varied aspects and briefly describe the existing literature in each of the aspects systematically. 
\section{Taxonomy: Cloud-Fog-Edge System}
\begin{figure*}
\centering
\includegraphics[scale = 0.4]{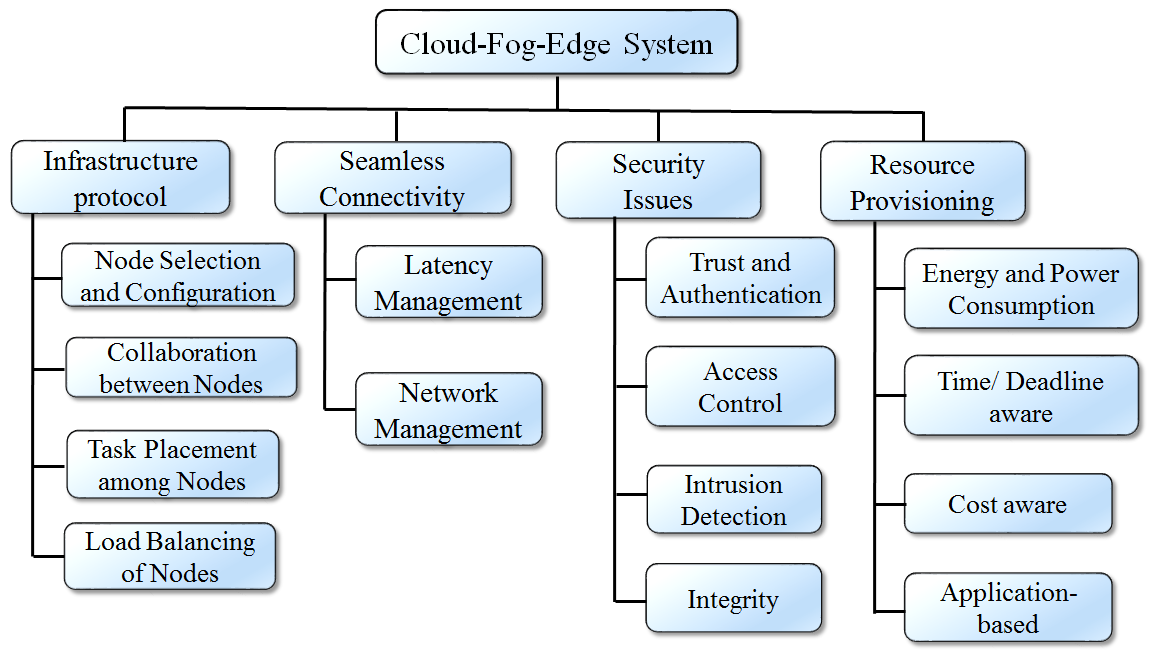}
  \caption{Taxonomy of Cloud-Fog-Edge system}
  \label{fig2} 
\end{figure*}

In this section, we discuss the taxonomy of Cloud-fog-edge collaborative system. For this taxonomy, we explore the system aspect and the associated challenges in this domain. Fig. \ref{fig2} shows the cloud-fog-edge system taxonomy, where we observe four broad aspects, such as, \textit{infrastructure protocol}, \textit{seamless connectivity}, \textit{security issues} and \textit{resource provisioning}. 

\subsection{Infrastructure protocol}
The cloud-fog-edge collaborative system has different challenges. Amongst them, few challenges are identified and discussed here. 
\par As mentioned earlier, varied devices based on the requirement of the system can act as fog/ edge nodes.  In general, the fog/ edge nodes are geographically-distributed. These are deployed at varied places, such as, shopping malls, roads, airport-terminals etc. These nodes are virtualized and have network connectivity along with storage and computation capabilities. Some of the works classify the nodes as micro/ macro/ nano-servers based on their physical size \cite{server}. Zeng et al. \cite{server2} utilize fog server as computational and storage servers in a software-defined embedded system. In \cite{mobiiost}, authors use RSU as fog-device to communicate between end-user and cloud-data center, since the user is in move. Cloud-based services are extended using cloudlets \cite{cloudlet}, \cite{cloudlet2}. The authors present a methodology for user-cloudlet association to reduce the cost in fog computing \cite{cloudlet2}. The conventional base stations are used for data signal processing and connectivity \cite{base}. Small cell base stations are also used as fog nodes. 
\par The communication or collaboration among the nodes is a important aspect as well. Fog nodes can form a cluster and collaborate among them for execution of a task \cite{scal}, \cite{load4}. These clusters are formed either the types of the nodes (homogeneity) or the location of the deployed nodes. Computational load balancing and functional sub-system
development have the higher priority while forming the cluster. Although, this cluster based technique is effective for some cases, the static clusters become the bottleneck in scalability of the system. In summary, based on the requirement (computation, storage and cost) of the application, proper node selection is very important.  Peer to Peer (P2P) collaboration among the nodes is another technique, which can be either hierarchical or flat order \cite{mobilefog}. There are several types based on proximity, such as home, local, non-local. Anyway, reliability and access control are the issues associated with P2P. Master-slave is another technique in this node-collaboration \cite{master}. 

\par Load balancing amongst these nodes are a challenging issue to prevent overloading of any particular node. A novel load balancing technique for edge data centers is presented in \cite{load1}. The authors in \cite{load2} propose dynamic edge selection to balance the loading of the nodes. Another work \cite{load3} explores the dynamic resource allocation and come up with a adaptive resource allocation scheme. The proposed method integrates utilization of bandwidth, computation resources by using dynamic load balancing. Oueis et al. \cite{load4} propose another dynamic load balancing technique by analysing the data flow in real-time. 
\par In task-placement. there is a trade-off between high capability infrastructure (cloud data centers) and connectivity issues. Several researchers have explored the cloud-edge topology \cite{task1},fog-to-cloud \cite{task2} or combined fog-cloud \cite{task3} system. The cloud-edge hybrid system for learning techniques has been proposed in \cite{task4}. The authors claim that the load variations in a neural network also should be considered for better efficacy. Each of the integrated system has its own advantages and disadvantages. It may be note that cloud servers have more computation and storage capabilities however, they are vulnerable to network connectivity lost issue. In such case, fog/ edge nodes exhibit less network communication cost and better latency. Therefore, the time-critical applications such as, emergency services or defence applications should adapt such system, where real-time response can be made. \subsection{Seamless Connectivity}
Seamless connectivity is a critical issue in enhancing the quality of service (QoS). Latency management consists of managing the service delivery
time by an accepted temporal threshold. The temporal threshold is measured by the maximum latency of a service request or the requirement regarding QoS. Some researchers have emphasized on efficient collaboration technique such that execution can be made faster \cite{load4}. Zeng et al. \cite{server2} minimizes the computation and communication latency by task distribution. A low-latency Fog network has been presented in \cite{con2} for better latency management. 
The aim of all these works to select the node capable to provide service in minimum delay.  
\subsection{Security Issues}
Security is a major concern as there are several communication between underlying network and cloud data-centers \cite{sec0}, \cite{sec00}.
\par Users authentication is one of the major aspect in fog/ edge based systems. Here, the components follow ``pay as you go" model, therefore, there is restricted access. There are various methods, like, \textit{user authentication}, \textit{device authentication}, \textit{data migration authentication and instance authentication} \cite{sec1}. Since the data are collected from end-users in most of the cases, proper privacy assurance is required \cite{scal}. 
\par The encryption in the fog/ edge nodes is also required, since the data is send to cloud datacenters from fog/ edge devices. Aazam et al. \cite{sec2}  has appended a data encryption layer in their system architecture for encryption. As fog/ edge nodes have limited resources, it is difficult to manage large concurrent service-requests. Here, \textit{Denial-of-Service (DoS)} is critical since it affects the system throughput at a large margin. Intrusion detection method is required to prevent such DoS attack. A work \cite{sec7} propose a cloudlet mesh based security framework to detect such intrusion on cloud and any intermediate communication. Access control is a reliable method to preserve the security and privacy of user. A fine-grained data access control scheme on attribute-based encryption (ABE) is presented in \cite{sec5}. Another Work \cite{sec6} proposes a policy-based resource access control in fog computing for secure interoperability. %
\par Integrity is another important part of privacy. A Lightweight Privacy-preserving
Data Aggregation (LPDA) scheme is proposed in \cite{sec11} where it can aggregate the data of hybrid IoT devices, and prevent any false data injection. It has also outperformed other existing approaches in terms of computational costs and communication overhead. A differential privacy-based query model is presented in \cite{sec9} for sustainable fog computing, where \textit{Laplacian mechanism} is utilized. This method has better efficiency and reduced energy consumption compared to existing methods. Huo et al. \cite{sec10} present a location difference-based proximity detection (LoDPD) protocol, where \textit{Paillier encryption algorithm and decision-tree theory} is used. 
\par Authentication is not sufficient because the devices itself are vulnerable to malicious attacks.
Trust is important factor here, where fog nodes verify the requests from the end-users or IoT devices, as well as, the end-users verify the services received from trusted fog/ edge nodes \cite{sec8}. In other words, system needs to confirm whether the fog/ edge nodes are secured, thus, a robust trust model is required. There are several issues like how to measure trust in the fog/ edge nodes and which attributes should be included in the trust model. 
The conventional trust models in cloud computing are not useful due to lack of centralized
management and mobility issues. \par However, there are unsolved challenges like how to implement intrusion detection in geo-distributed, large-scale, high-mobility fog computing system to satisfy latency requirement. Further studies need to investigate how fog computing can be beneficial for intrusion detection on both client side and the centralized cloud side. 

\subsection{Resource Provisioning}
Another challenge is to efficiently allocate cloud/fog/edge computing infrastructure to different services. At each time-instance, IoT device or end-users can request huge number of service-request, but each fog/ edge device is resource-constraint. Therefore, the components (edge and fog device/ node) should be efficiently managed. 
The resource management among fog/edge
nodes is another aspect here. This should be considered based on service-requirements and service-availability, energy consumption. In summary, the mapping of the resources to fog/edge service nodes is compelling
issue. 
\par Since the fog/ edge nodes have limited computing and storing resources, it is not possible always to satisfy all service-requests. To resolve this, \textit{satisfaction function} is formulated to measure the allocated resources to execute the service-request. The satisfaction function is defined as \cite{sur1}: 
\begin{equation}
    f(res)=\left\{\begin{matrix}
 \log(res+1) & 0\leq res \leq res_{min} \\ 
 \log(res_{max}+1) &  res \geq res_{max}
\end{matrix}\right.
\end{equation}
where $f$ is the satisfaction function, the allocated resource and maximum resource are denoted by $res$ and $res_{max}$ respectively. The objective is to maximize the overall $f(res)$ for all end-users is defined as: 
\begin{equation}
    \textbf{Objective} \; \;  \max{f_{All}} 
\end{equation}
\begin{equation}
    \textbf{s.t.} \left\{\begin{matrix}
f_{All} = \sum_{u=1}^{U}\{pr_u \times f_u(res_u)\}\\ 
res_1 + res_2 + \dots + res_u \leq RES\\ 
pr_1+pr_2+ \dots + pr_u = 1 \\
res_1, res_2, \dots, res_u \geq 0
\end{matrix}\right.
\end{equation}
where the priority value, user and total resource are defined as $pr$, $u$ and $RES$ respectively \cite{sur1}.
\par A simulation toolkit for measuring the efficacy of any fog-based framework is presented in \cite{ifogsim}. A \textit{LP-based two-phase heuristic algorithm} resource management framework is proposed in \cite{res2} in fog-based medical cyber-physical system. 
\section{Taxonomy: Mobility Management}
\begin{figure*}
\centering
\includegraphics[scale = 0.4]{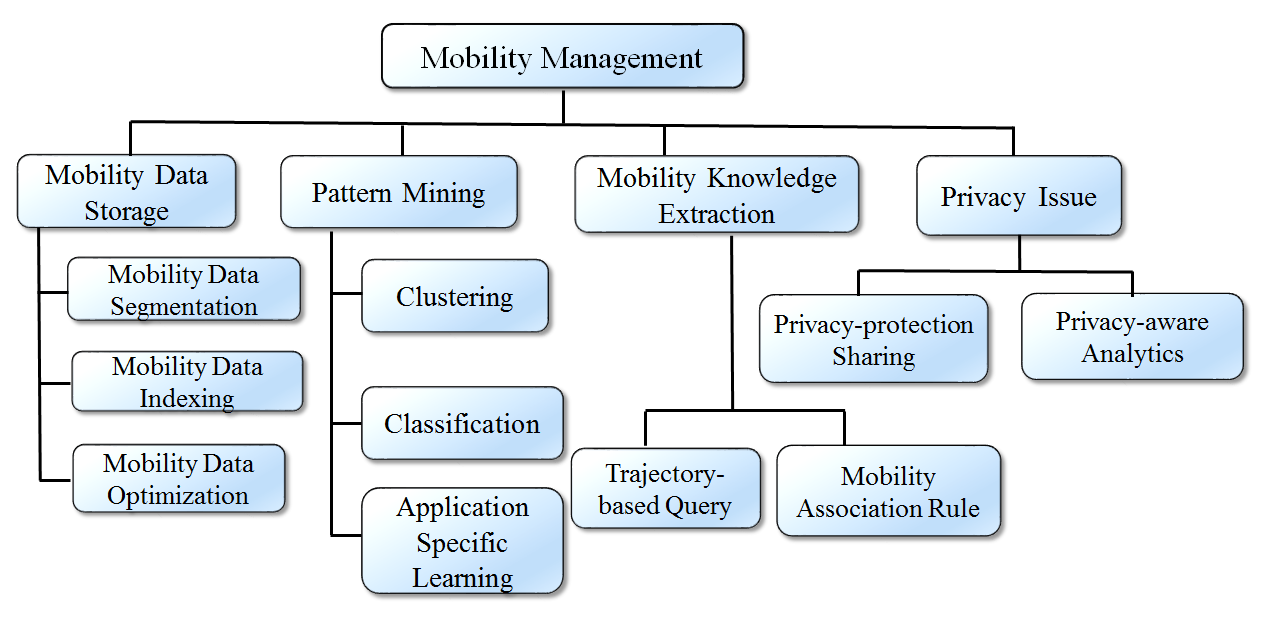}
  \caption{Taxonomy of Mobility-aware system}
  \label{fig3} 
\end{figure*}

With the advancement of Global Position Systems (GPS) and location acquisition technology, there is a growing need to analyse the huge amount of accumulated GPS log. The time-stamp location traces (latitude, longitude) is defined as \textit{trajectory}. Several researches have been carried out in this domain. After carefully studying the existing literature, we come up with the taxonomy (refer Fig. \ref{fig3}) of mobility-aware system where \textit{mobility data storage}, \textit{pattern mining}, \textit{mobility knowledge extraction} and \textit{privacy issues} are highlighted. 
\par Trajectory database size is huge due to its dynamic nature. There are several works on efficient trajectory data storage technique. This includes mobility data segmentation, mobility data indexing and trajectory data optimization. 
\par There are varied trajectory indexing techniques. In \cite{hr}, a multi-version structure, named \textit{HR+}, is proposed, where a node can store different time-stamp entries. Thus it reduces the space complexity. A two-level index structure is proposed in \cite{chakka}, where the index of spatial and temporal information is decoupled. Ghosh et al. propose a \textit{k-level temporal hash-based scheme} \cite{mario}, where a hash function is used to store the movement data efficiently. Another work, named \textit{Trajstore} \cite{trajstore} dynamically co-locates and compresses data on same disk by creating an optimal index. Zhou et al. \cite{grid} presents a grid-based index. Here, the study area/ region is segregated into different rectangular cells of fixed size. In each such segment, the information of the mobility information is stored separately. Another work \textit{Mobi-IoST} \cite{mobiiost} presents an unique grid-based trajectory information segmentation, where in each segment a fog node is present and the fog node stores such information. 
\par Pattern mining is another major aspect of trajectory data analysis. The objective is to \textit{find the intent behind any move} \cite{thump}, and making sense of the trajectory log \cite{sense}. In this context, \textit{semantic trajectory} is defined, where raw trajectory log is complemented with additional information such as, point-of-interests (POI) of the path followed, stay-points and duration, speed, transportation mode etc. The process of appending such semantic information is defined as \textit{semantic enrichment}. Different works have proposed several methods to semantically enrich the movement log based on the requirements of the application. In \textit{Traj-cloud} \cite{trajcloud}, the authors describe the process of geo-tagging the trajectory points using reverse geo-coding. Google place API is used to extract the POI-information. Another work, named \textit{STMaker} \cite{make} presents a system, where raw
GPS log is segmented based on the behavioral features of the moving agent. Then, each such segment is identified by a short textual description. Furthermore, short textual messages from social networking sites also append semantic information about the movement. Based on this idea, TOPTRAC \cite{sem2} detects latent topic in trajectory dataset. It also extracts mobility patterns among different semantically connected regions. A clustering-based approach to find out the semantic regions is proposed in \cite{sem3}. 
\par Clustering and classification are another important aspects of mobility pattern mining. Trajectory data is presented as sequences of \textit{stop} and \textit{move} along with a temporal scale. Clustering is beneficial for grouping the similar type of movements. Thus, trajectory clustering techniques are presented by a number of works. Amongst varied distance-based clustering, EDR (Edit Distance on Real sequence), DTW (Dynamic Time Warping) and LCSS (Longest Common Subsequences) are popular. A partition-and-group based method is presented in \cite{clust1} to extract common trajectory segments. This TRACLUS framework use the minimum description length (MDL) and a density-based line-segment clustering method. Representative
Trajectory Tree is proposed in \cite{clust2} for temporal-constrained sub-trajectory clustering. A novel trajectory clustering approach using
deep representation learning is presented in \cite{clust3}, where a sequence to sequence auto-encoder is utilized. TULER presents a RNN based model to extract the dependency of checkins \cite{clust4}. The work \textit{MovCloud} \cite{movcloud} proposed a mobility-clustering algorithm based on the semantic behaviour of the users and the clustering algorithm is deployed in the cloud servers for fast execution. Classification of trajectories aim to train the model and use it for prediction. A work \cite{class1} augments duration information to enhance the prediction accuracy, along with spatial distribution and shape of the trajectories as features of the classification algorithm. Another work by Ghosh et al. \cite{class2} categorizes users based on their regular movement pattern. The study has been carried out in an academic campus, and the algorithm can classify users as \textit{professor}, \textit{student}, \textit{staff} categories effectively.
\par To retrieve information from a dataset, query processing is important. There are varied types of queries, namely, \textit{point query}, \textit{range query}, \textit{trajectory query} etc. The range or R-query ($RangeQ(S, T)$) finds all trajectory segments which intersects the given spatial (S) and temporal (T) extent.
\begin{equation*}
    RangeQ(S,T) \rightarrow Traj
\end{equation*}
where $Traj$ is the set of trajectory segments within $S$ spatial and $T$ temporal extent. 
The T-Query or trajectory-based query finds all trajectory segments of a moving agent (a) within the temporal interval
(T).
\begin{equation*}
    TrajectoryQ(a,T) \rightarrow Traj
\end{equation*}
Here, $Traj$ is the output trajectory of the query. Several researchers have deployed novel methods to resolve queries effectively. A location based searching is proposed in \cite{29}, where different locations are assigned different importance or priority. For instance, location with geo-tagged information (such as, photograph) is more important than others. It finds the k-most important connected trajectories. Range queries are studied in \cite{30}, \cite{31}. Vieira et al. \cite{35} defines \textit{pattern query} where trajectory segments with specific movement features are extracted. Aggregated queries \cite{18} produces an aggregate measure. There are also other context-based or application-based queries \cite{36}, \cite{37}. Mobility association rule defines the interrelation of two or more mobility events in temporal scale \cite{mario}, \cite{assoshr}, \cite{ghosh2017exploring}. 
\section{Taxonomy: Location-aware Services}
\begin{figure*}
\centering
\includegraphics[scale = 0.4]{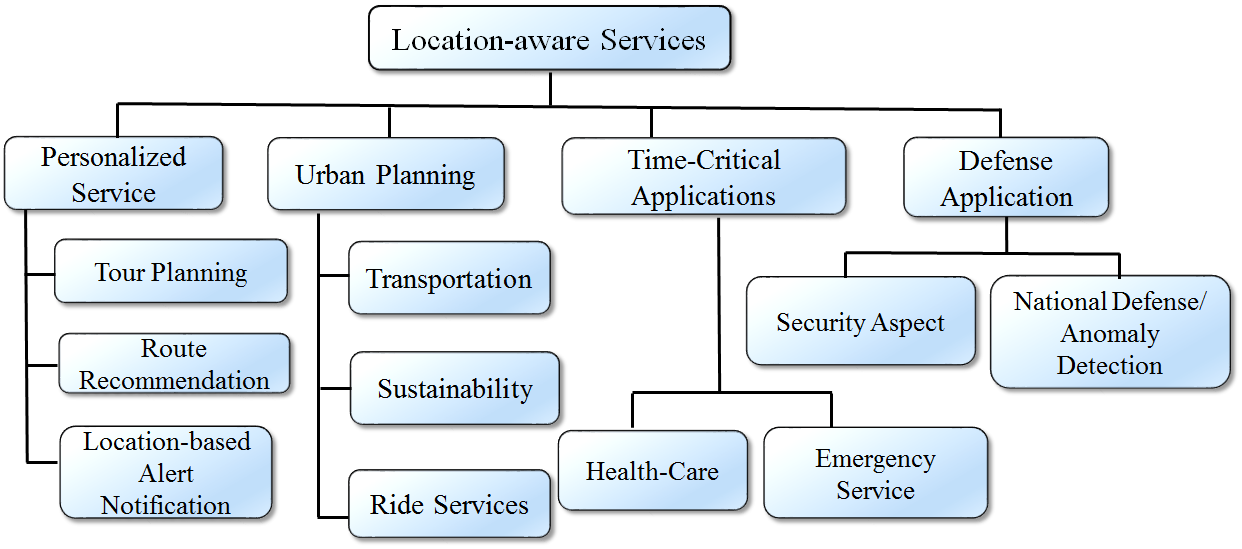}
  \caption{Taxonomy of Location-aware Services}
  \label{fig4} 
\end{figure*}

Location-aware services utilize the geographical location information to provide services to the end-users. Fig. \ref{fig4} illustrates different types of location-based services, namely, \textit{personalized service}, \textit{urban planning}, \textit{time-critical applications} and \textit{defence applications}. 
\begin{table*}[!htb]
\centering 
    \caption{Classification of Location-based Applications and Approaches}
    \resizebox{\textwidth}{!}{%
    \begin{tabular}{|r|l|l|l|}
  \hline
  Type of Service & Author and year & Application & Approach  \\ \hline
   & Han Su et al., 2019 \cite{ap1} & Personalized route description & User knowledge measurement and \\
& & system & route summarization \\ \cline{2-4} & Jun Suzuki et al., 2020 \cite{ap2} & Assign Personalized visited & Visited POI selection based and \\
& & points & 0-1 ILP formulation  \\ \cline{2-4}
& Zhao et al., 2020 \cite{ap3} & Personalized location & Sentimental-spatial \\
& & recommendation & POI mining  \\ \cline{2-4}
  Personalized  & Ghosh et al., 2018 \cite{ap4} & Location prediction from & Hierarchical and layered Hidden \\
  Service & & sparse trajectory data & Markov model (HMM) construction \\ \cline{2-4}
& Tarik Taleb et al., 2017 \cite{ap5} & Network slicing for  & Mobile network personalization service \\
  & & personalized 5G mobile  & orchestrator (MNP-SO) and the \\
& & telecommunication & mobile service personalization service \\ \cline{2-4}
& Ghosh et al., 2018 \cite{activity} & Activity-based user profiling & Allens' temporal calculus based\\
  & & & activity data analysis \\ \cline{2-4}
  & Han Zou et al., 2019 \cite{ap7} &  Inferring User identity  & WiFi-enabled nonintrusive device and \\ 
  & & and mobility & user association scheme \\ \cline{2-4}
  & Fei Wu et al., 2016 \cite{ap8} & Personalized annotation of & Markov random field to \\
  & & mobility records & maximize the consistency \\ \hline \hline 
  
   & Kong X et al., 2017 \cite{taxi1} &  Recommendation of services  & TLR model based on Gaussian process  \\ 
  & & to taxi drivers & regression and statistical approaches  \\ \cline{2-4}
  & Boting Qu et al., 2019 \cite{taxi3} & Profitable taxi travel &  Probabilistic network model \\
  & & route recommendation & and Kalman filtering \\ \cline{2-4} 
  &  Gang Pan et al., 2012 \cite{taxi7} & land-use classification & Support Vector Machine (SVM) classifier \\
  & & from taxi trajectory data & with features extracted \\ \cline{2-4}
 Urban & Hua Cai et al., 2019 \cite{taxi8} & Environmental benefits of & Quantifies the environmental benefits \\
  Planning & & taxi ride-sharing & of taxi ride sharing \\ \cline{2-4}
  & Tingting Li et al., 2019 \cite{taxi4} & Emission pattern mining & Spatial and temporal dynamic  \\
  & & for pollution detection &  emission patterns in varied traffic zones\\ \cline{2-4}
 &  Gong et al., 2016 \cite{taxi2} & Inferring trip-purposes from & Spatio-temporal analysis and \\
 & & taxi trajectory data & probability modelling by Bayes' rules \\ \cline{2-4} 
 & M Ota et al., 2017 \cite{taxi5} & Simulation of taxi ride-sharing & Linear optimization algorithm \\ 
 & & & and efficient indexing scheme \\ \cline{2-4} 
 & SP Chuah et al., 2016 \cite{taxi6} & Designing and optimization & Clustering of taxi-rides and  \\
 & & bus-routes & optimization problem to design bus-routes \\ \hline \hline 
 Time-Critical & Ghosh et al., 2019 \cite{mobiiost} & Recommending optimal path for & Probabilisitic graphical model and \\
Applications  & & and actuating signals for emergency & k-order Markov chain \\ 
 & & services (say, health-care)  & \\ \cline{2-4}
 & Mukherjee et al., 2020 \cite{ioht} & IoHT for personalized health & Generative adversarial networks based \\
 & & monitoring and recommendation & analysis \\ \hline \hline
 
 Defence & Du Bowen et al., 2019 \cite{det2}, \cite{det1}  & Identifying pickpocket suspects & Two-step framework of regular \\
Applications  & & from check-in data & passenger filtering and  \\ 
 & & & suspect detection from movement traces\\  \hline \hline
 
\end{tabular}%
}
\label{tab1}
\end{table*} 

\par Personalized service includes the notification or recommendation sent to the user based on her location. For instance, user's location is nearest to a new shopping mall, which is providing discounts on specific items. The system can provide alert to the user regarding this. Again, the system can predict probable congestion on a road-segment and notifies the user apriori to avoid the road-segment. Urban planning consists of sustainable solution in terms of energy and power consumption and smart transportation. Location-based services can also benefit to time-critical applications, where real-time response is required. defence application can also get support from location-data analytics. 
\par There are huge number of works where several challenging applications are mentioned and mobility analytics or location-based data mining supports these applications. Table 1 depicts few of these applications and the approaches followed in those works. 
The work by Han Su et al. \cite{ap1} explores how to make the route description more customized and intuitive depending on the user. The authors present the problem of extracting optimal partition of a given route that maximizes the familiarity of user and generates proper sequence. The visited POI-assignment task is carried out in \cite{ap2} where authors formulate the problem using 0-1 ILP formulation. It may be noted that characteristics and attributes of geographical locations are important for location recommendation. Zhao et al. \cite{ap3} proposes a novel method of personalized location recommendation by sentimental–spatial POI mining (SPM). The incomplete or sparse data is a major problem in predicting next location sequences efficiently. The authors in \cite{ap4} propose a novel layered and hierarchical HMM with other contextual data. The authors in \cite{ap8} aim to annotate the mobility records by their semantics (what the user is doing at that location). The work does not assume the availability of training data. Similarly there are works on sustainable polices like reducing carbon footprint by ride-sharing system \cite{taxi8} \cite{taxi4}. Another interesting application is presented in \cite{det1} and \cite{det2}, where the authors find out pickpocket suspects from large scale transit data. 
\section{Conclusions and Future Research Directions}
With the proliferation of GPS-enabled devices and smart hand-held devices, huge volume of data is generated. This huge amount of data can be beneficial for mining behavioural patterns of users and thus fostering challenging applications in our daily life. However, analysing such big amount of data is not possible for resource-constraint devices. Therefore, cloud technology is important. But, communication with distant cloud servers may affect the latency of the service-response and reduces the QoS. Therefore, partial computation must be carried out at the edge of the network, and thus, fog/edge nodes are incorporated. In this book chapter, we provide a brief summarization of issues and challenges in a mobility-driven cloud-fog-edge based framework for facilitating location-based services. We have presented three taxonomies of existing works in system aspect, mobility management and several types of location-based services. Further, a tabular representation of few prominent applications have been highlighted. Although there are huge amount of works in this domain, we have identified few challenges and opportunites which can be explored in the future. 
\begin{itemize}
    \item \textbf{Heterogeneity of the layers:} There is a big challenge in the heterogeneous nature of the components at fog/ edge layers. 
    The system should be able to orchestrate several different types of devices with heterogeneous cores. Significant architectural advancements should be made both from hardware and deployment of such system. 
    \item \textbf{Security:} Although there are few works in security mechanism, but the cross-layer security or privacy policies are still un-explored and a big threat to both system and end-users.
    \item \textbf{Sustainable mobility:} Increasing power and energy consumption and thus increasing carbon footprint are big issues in present times. Although there are few works on ride-sharing to reduce the carbon footprints, considering budget
and deadline \cite{das2021lyric}, in reality this is still a big issue. The proper mechanism to deploy such initiatives is yet to be done. There is also concern of user-security while sharing rides with unknown people. 
    \item \textbf{Sharing data with people:} Sharing data (transportation, urban planning, healthcare facilities etc.) with citizen is absolutely necessary. This not only brings transparency, people can make proper decision if they know the available resources. While few developed countries have systematic data sharing policies, developing countries like India, is still far behind. A significant progress can be made if the research community can come up with appropriate data sharing policies and modules for such cases. 
\end{itemize}

We believe that the book-chapter will provide a brief but comprehensive review of cloud-fog-edge collaborative framework to the readers. 

\section*{Acknowledgment} This work is partially supported by TCS PhD research fellowship.

\bibliographystyle{unsrt}
\bibliography{biblio}
\end{document}